\documentclass[%
 reprint,
 amsmath,amssymb,
 aps,
]{revtex4-2}

\usepackage{graphicx}
\usepackage{dcolumn}
\usepackage{bm}

\renewcommand{\d}{\text{d}}

\begin{document}

\preprint{APS/123-QED}

\title{Sensitive dependence of the linewidth enhancement factor on electronic quantum effects in quantum cascade lasers}

\author{Martin Francki\'e}
\author{Mathieu Bertrand}
\author{J\'er\^ome Faist}
 \affiliation{Institute for quantum electronics, ETH Zürich, Auguste-Piccard-Hof 1, 8093 Zürich, Switzerland.}

\date{\today}

\begin{abstract}

The linewidth enhancement factor (LEF) describes the coupling between amplitude and phase fluctuations in a semiconductor laser, and has recently been shown to be a crucial component for frequency comb formation in addition to linewidth broadening. It necessarily arises from causality, as famously formulated by the Kramers-Kronig relation, in media with non-trivial dependence of the susceptibility on intensity variations. 
While thermal contributions are typically slow, and thus can often be excluded by suitably designing the dynamics of an experiment, the many quantum contributions are harder to separate. In order to understand and, ultimately, design the LEF to suitable values for frequency comb formation, soliton generation, or narrow laser linewidth, it is therefore important to systematically model all these effects. 
In this comprehensive work, we introduce a general scheme for computing the LEF, which we employ with a non-equilibrium Green's function model. This direct method, based on simulating the system response under varying optical intensity, and extracting the dependence of the susceptibility to intensity fluctuations, can include all relevant electronic effects and predicts the LEF of an operating quantum cascade laser to be in the range of 0.1 - 1, depending on laser bias and frequency. We also confirm that many-body effects, off-resonant transitions, dispersive (Bloch) gain, counter-rotating terms, intensity-dependent transition energy, and precise subband distributions all significantly contribute and are important for accurate simulations of the LEF.
\end{abstract}

\maketitle


\section{\label{sec:introduction}Introduction}

In a semiconductor laser, current injection modifies the carrier distribution, changing the imaginary part of the susceptibility - i.e. the  gain necessary for laser operation. As expressed by the Kramers-Kronig relations, the real and imaginary parts of the susceptibility are however coupled, and the ratio of their change upon population injection $\delta n$  is quantified by the linewdith enhancement factor
    \begin{equation}
       \alpha = \frac{\partial \chi'/\partial \delta n}{\partial \chi" /\partial \delta n}
   \end{equation}
   This parameter was introduced by C.H. Henry in his celebrated paper\cite{henry_theory_1982} to explain the enhancement of the linewdith of these devices beyond the Schawlow-Townes limit\cite{schawlow_infrared_1958}. Indeed, changes of the refractive index accompanying the fluctuation in carrier density from the spontaneous emission during laser operation act as a phase modulation of the laser emission, thus further broadening the linewdith by a factor of $1+\alpha^2$.  However, this coupling between gain and refractive index, is not limited in its effects to the laser linewidth but is also fundamental in many other aspects of the laser dynamics such as frequency chirp or the effect of optical feedback\cite{lang_external_1980}\cite{saito_oscillation_1982}. It was also recently realized that the linewidth enhancement factor played a key role in formation of optical frequency combs\cite{silvestri_coherent_2020} and 
   solitons\cite{franckie_self-starting_2022} in media with fast gain saturation as it can be seen, combined with gain saturation, as an optical Kerr effect.
   
   In the last 20 years, the quantum cascade laser (QCL) has emerged as a powerful, compact, and versatile source of coherent mid-IR radiations with a wide range of operation between 3$\mu$m and 16$\mu$m of wavelength covering the molecular fingerprint region of gases. As QCLs exhibit an atomic-like joint density of state, it was immediately noted that the value of their linewidth enhancement factor should be very small if not vanishing\cite{faist_quantum_2013}. Indeed, measurements using a high frequency modulation of a single frequency device yielded a small albeit non-zero value~\cite{aellen_direct_2006}. One source of confusion has been the fact that thermal effects, larger in QCLs because of their high dissipation, is also responsible for changes in the refractive index and lead to very significant contributions to the linewidth enhancement factor at low frequencies~\cite{hangauer_high_2014}.
   
 The physical origin of the LEF in a QCL is a nontrivial combination of both macroscopic and microscopic effects. While some of these occur on different time scales, they can all significantly contribute to the LEF. Broadly, these effects can be divided into those originating from the electronic system and from the surrounding environment. Gaining detailed information on the former is experimentally very challenging. In order to control the LEF of such devices it is therefore necessary to understand the light-matter interaction via microscopic modeling.

\begin{figure}
    \centering
    \includegraphics[width=\linewidth]{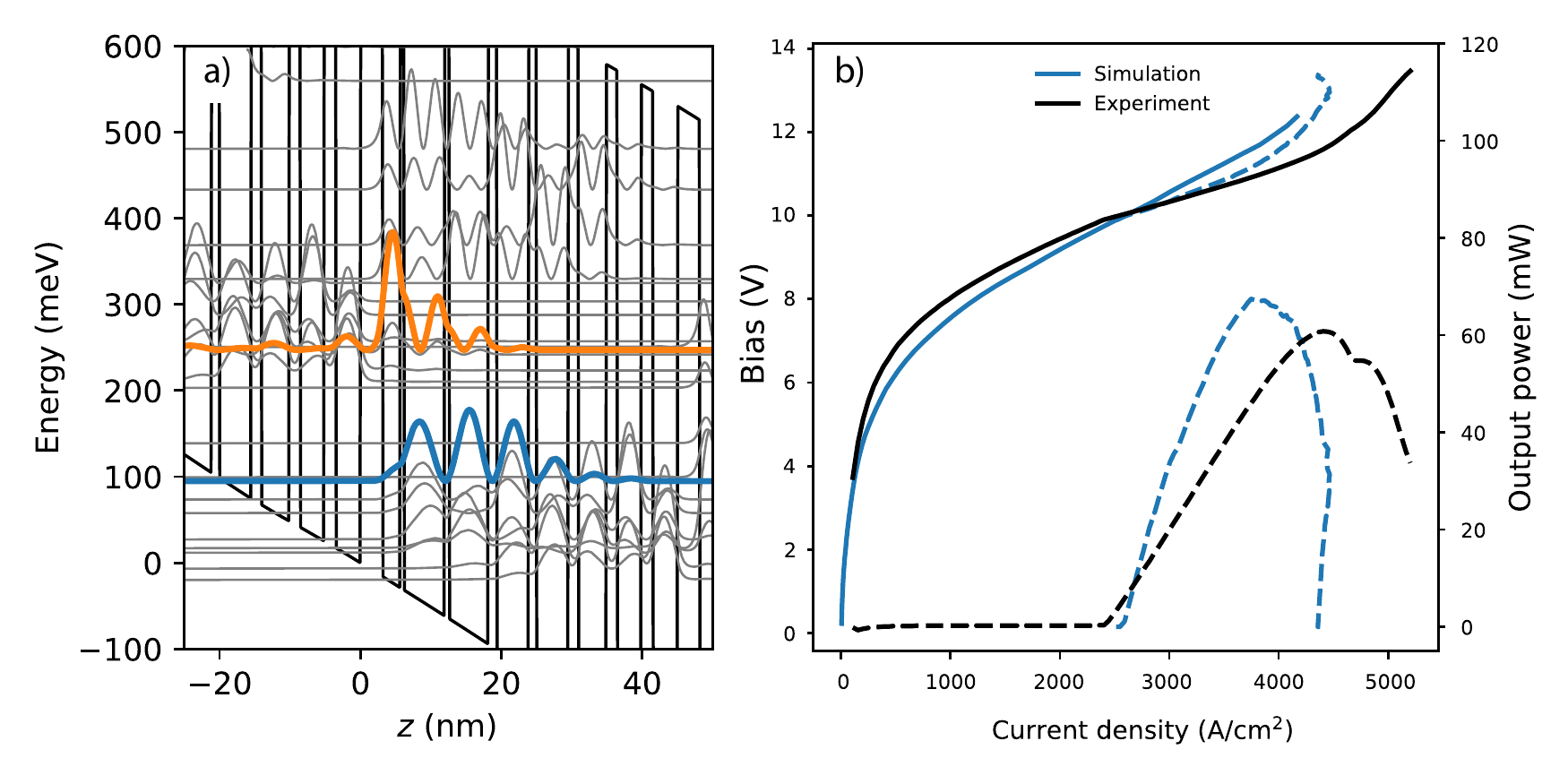}
    \caption{NEGF simulations of the mid-infrared QCL EV2016\cite{wolf_quantum_2017}. (a) Conduction band structure and the electron wavefunctions at an applied bias of 10 V. The upper (orange) and lower (blue) laser states are highlighted. (b) Light-current-voltage (LIV) curves obtained from experimental measurements of a DFB QCL at -20$^\circ$C heat sink temperature, as well as NEGF simulations. The dashed line corresponding to the NEGF current density shows the photo-driven current, while the solid one shows the IV without lasing. The lattice temperature of the simulations is bias-dependent, and has been extracted from the experiment by measuring the temperature-dependent frequency shift of the DFB emission.}
    \label{fig:LIV}
\end{figure}

The LEF in mid-IR QCLs has been modelled for a reduced two- or three-level system using density matrix models\cite{opacak_frequency_2021, wang_rate_2018, liu_importance_2013}, which are capable of capturing the coherences induced by the light-matter interaction, from which the complex susceptibility can be calculated. Using a more general technique, Pereira studied the LEF of a three-level quantum well system using nonequilibrium Green's function (NEGF) theory\cite{pereira_linewidth_2016}, which includes many-body quantum effects such as carrier-carrier interactions and self-energy (Lamb) shifts, and counter-rotating terms. As mentioned above, also other non-resonant transitions play a crucial role for correctly modeling the LEF\cite{jungho_kim_theoretical_2004}. In these previous studies, several important effects have been identified and studied separately, namely dispersive gain\cite{wacker_coexistence_2007, willenberg_intersubband_2003}, counter-rotating terms\cite{pereira_intersubband_2011}, non-parabolicity\cite{liu_importance_2013}, and the contribution from non-resonant transitions.\cite{jungho_kim_theoretical_2004} However, until now no model including all these effects have been applied to the calculation of the LEF in a QCL, and thus the relative importance of these effects is still not known.

In order to accurately model the LEF, we utilize an NEGF model\cite{wacker_nonequilibrium_2013}, which accounts for non-thermal subband distributions, non-parabolicity\cite{lindskog_injection_2013}, and many-body quantum effects. Since it includes all relevant active region electronic quantum states in a basis-invariant scheme, non-resonant contributions are accurately taken into account, even when states are close to resonance. In addition, NEGF theory goes beyond first order perturbation theory for scattering as well as light-matter interaction, although the optical response is currently limited in our model to a single frequency at a time, which is not a limiting factor for the present study. The accuracy of the model with respect to the experimental device is benchmarked in Fig.~\ref{fig:LIV}, where the threshold current has been fitted by adjusting the interface roughness parameters within the experimental uncertainty limits. It is clear that the model can reproduce the output power and the transport well for these parameters, which means that the nonlinear optical susceptibility can be obtained with reasonable accuracy.

\section{Nonequilibrium Green's function simulations}

\begin{figure}
    \centering
    \includegraphics[width=\linewidth]{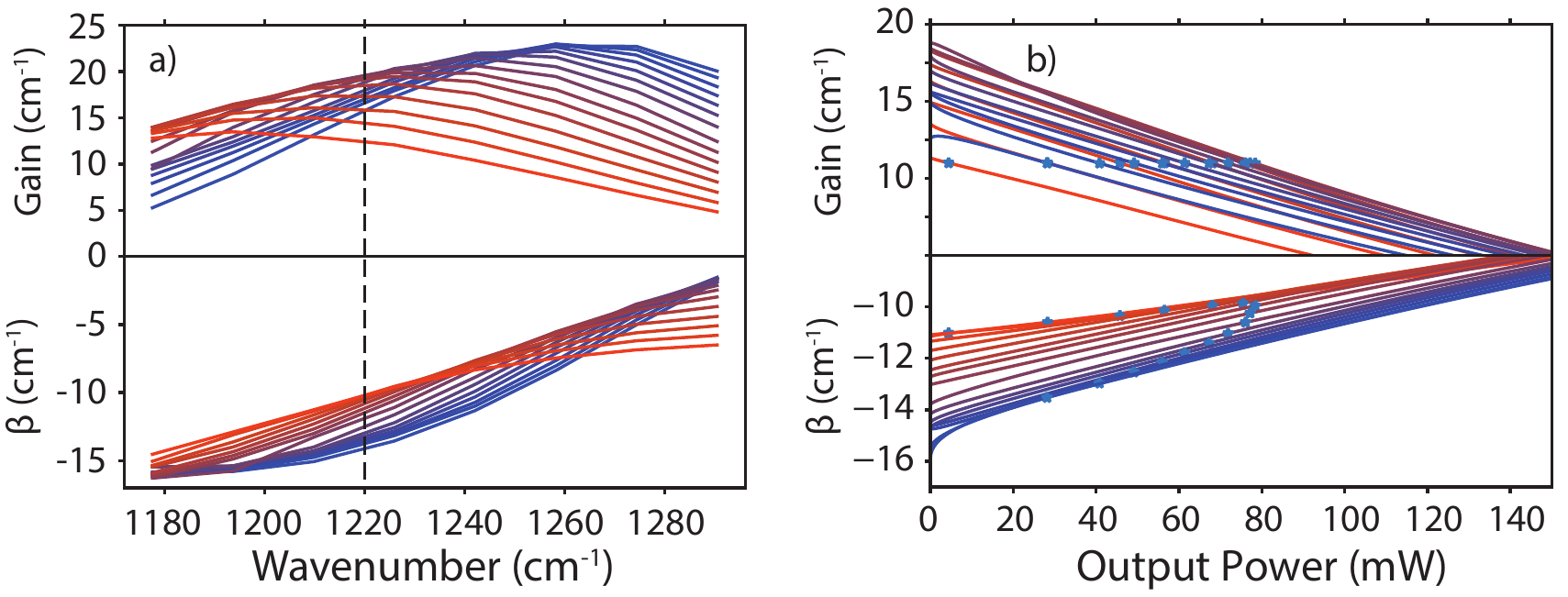}
    \caption{NEGF simulations of the gain and dispersion $\beta$, in (a) as functions of bias (from 220-290 mV/period in steps of 5 mV/period) and vanishing optical power and in (b) as functions of output power, assuming a facet reflectivity of 30\% for the biases in (a). In (b), the frequency has been fixed to $\hbar\omega = 150$ meV $= 1210$ cm$^{-1}$, marked by the dashed line in (a).}
    \label{fig:GainSat}
\end{figure}

\begin{figure}
    \centering
    \includegraphics[width = 0.75\linewidth]{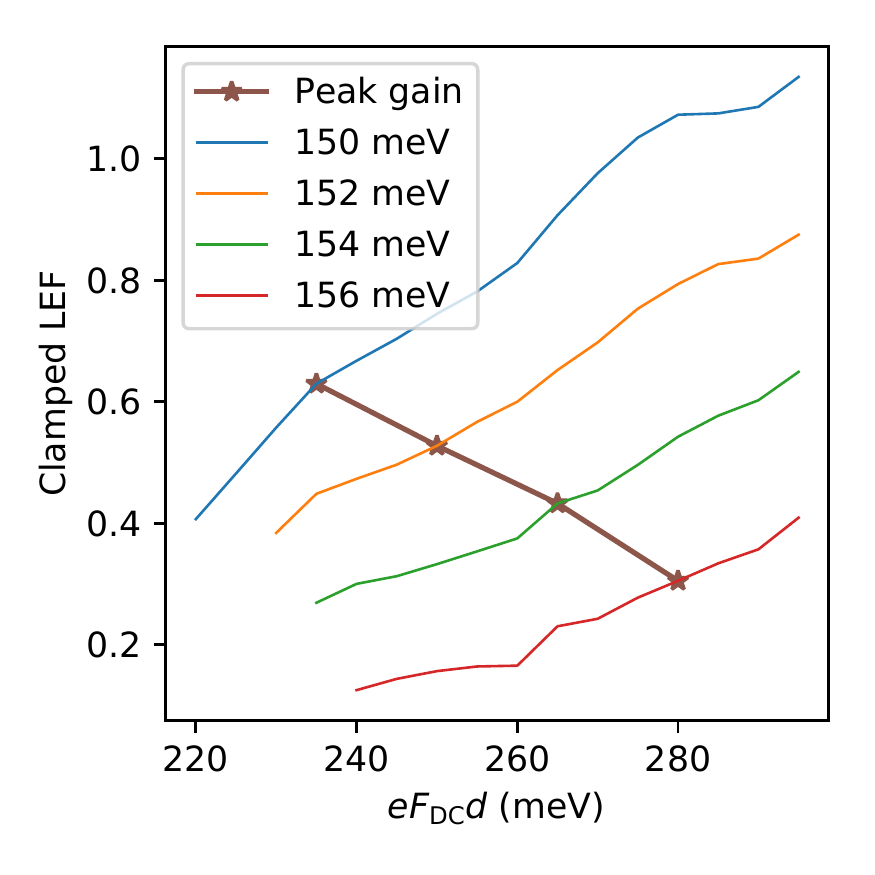}
    \caption{Linewidth enhancement factor at the gain clamping AC field, as a function of bias. Solid lines show traces for four different frequencies, while the asterisks show the values evaluated at the peak of the gain curve.}
    \label{fig:LEF_bias}
\end{figure}

The main object to calculate is the dynamical lesser Green's function, which is found through a Fourier expansion in harmonics $h$ of the laser frequency as
$G^<_{mn}(\mathbf{k}, E, t)\equiv\sum_h G^<_{mn,h}(\mathbf{k}, E)e^{-\text{i}h\omega t}$.\cite{wacker_nonequilibrium_2013} Relevant dynamical quantities can then be found via the higher-order response of $G^<$, such as the current density
\begin{eqnarray}
    &&J(z) = \sum_h J_h e^{-\text{i}h\omega t} = \nonumber \\ 
    && -\frac{\text{i}}{A}\sum_\mathbf{k}\sum_{mn} J_{mn}(z) \int \frac{\text{d} E}{2\pi}\sum_h G_{mn, h}^<(\mathbf{k}, E)e^{-\text{i}h\omega t}.
\end{eqnarray}
Here, $A$ is the device area, $J_{mn}$ are matrix elements of the current operator \cite{lindskog_injection_2013}, $E$ and $\mathbf{k}$ are the energy and momentum, respectively, and $h$ is the Fourier expansion coefficient. The intensity gain $g$ and the active region contribution to the propagation constant $\beta$ are, respectively
\begin{eqnarray}
g(\omega) = -\frac{J_1 + J_{-1}}{c\varepsilon_0\sqrt{\varepsilon_r}F_\text{AC}}, \\
\beta(\omega) = -\frac{J_1 - J_{-1}}{2c\varepsilon_0\sqrt{\varepsilon_r}F_\text{AC}},
\end{eqnarray}
where $F_\text{AC}$ denotes the AC field strength and $\Delta k = \beta - \text{i}\frac{g}{2}$ is the complex wavevector due to the active medium. For each value of the AC field strength, the steady-state values for the different harmonics $G_{mn,h}^<$ and lesser self-energies $\Sigma^<_{mn,h}(\mathbf{k},E)$ up to a certain truncation order $N_h$, are converged. The required $N_h$ depends on the frequency and the field strength required to reach gain clamping at a certain bias voltage. In these simulations $N_h \leq 2$ was used. 
\\\\
The laser is modelled by sweeping the intracavity intensity for each applied DC bias to find the gain clamping condition shown by asterisks in Fig.~\ref{fig:GainSat}. This allows the computation of the photo-driven current and output power of the operating laser\cite{lindskog_comparative_2014} shown in Fig.~\ref{fig:LIV}. At these operation points, the effect of a small change in the intracavity intensity on $g$ and $\beta$ is then investigated, in order to deduce the LEF using the relation (using $\frac{d}{d\delta n} = \frac{d \delta n}{d I}\frac{d}{dI}$)
\begin{equation}
    \alpha = \frac{\mathcal{R}\{\d \tilde{n}(I)/\d I\}}{\mathcal{I}\{\d \tilde{n}(I) / \d I\}} 
    = -2\frac{\partial \beta/\partial I}{\partial g/\partial I},
    \label{eq:alpha}
\end{equation}
where $\tilde{n} = \frac{c}{\omega}(\beta - \text{i}\frac{g}{2})$ is the complex intensity-dependent refractive index. The resulting LEF at operating conditions, assuming a threshold gain of $g_\text{th} \approx 11$ cm$^{-1}$ calculated from the threshold current of devices of different lengths, is shown in Fig.~\ref{fig:LEF_bias}. While a free-running QCL operates in the single-mode regime at the frequency of peak gain $\omega_\text{max}$, a QCL operating as a frequency comb or a single-mode distributed feed-back (DFB) laser can host optical frequencies also away from the gain peak. Furthermore, since $\omega_\text{max}$ changes with bias, it is important to consider the LEF as a function of the emission frequency as highlighted in Fig.~\ref{fig:LEF_bias}. The solid lines show the evolution of the LEF at particular frequencies, at the  alternatinc current (AC) field strengths at gain calmping, corresponding to DFB QCLs centered at those frequencies. In this case, the LEF increase linearly with bias. As the slope $\partial g/\partial I$ is nearly constant with bias (see Fig.~\ref{fig:GainSat} b), this is a result mainly of the increasing slope of $\beta$. We also show the points at the frequency of the gain peak at four different biases in Fig.~\ref{fig:LEF_bias}.  First we evaluate the LEF at a single frequency at a time, corresponding to a DFB laser operating at approximately the same frequency irrespective of bias, or the modes of a frequency comb.  When instead evaluating the LEF at $\omega_\text{max}$ for every bias point, the opposite trend is found, i.~e.~a decreasing LEF with bias. This rather non-intuitive result can be explained with the fact that, as the frequency increases, $\beta(\omega_\text{max})$ moves closer to the inflection point around 159 meV (see Fig.~\ref{fig:NEGF_DM}). This inflection point is expected from the real part of the susceptibility for a two-level system crossing zero, but is here shifted in frequency due to the presence non-resonant transitions. Therefore, the LEF decreases towards higher frequencies. Although the red shift of $\omega_0$ with intensity in Fig.~\ref{fig:NEGF_DM} a) is expected from dispersive gain\cite{willenberg_intersubband_2003}, the inflection point in $\beta$ is further from the peak gain, and shift of $\omega_\text{max}$ smaller, than expected from dispersive gain alone. Thus other effects, such as other transitions and intensity-dependent transition energy, also play a important roles. Since the LEF is defined as the ratio between the real and imaginary parts of $\d \chi/\d I$, the contributions to $\alpha$ are not additive. This means that failure to account for all these effects may lead to significant differences in the calculated LEF. In the following we will therefore quantify other contributions to the LEF.
\\\\
\section{Contributions to the LEF}

To lowest order, the main contribution to the LEF is a change in inversion with intensity, phenomenologically given by
\begin{eqnarray}
    \Delta N (I) = \frac{\Delta N(0)}{1 + I/I_\text{sat}}, \label{eq:saturation}
\end{eqnarray}
where $I_\text{sat}$ is the saturation intensity. In this case, only the value of the susceptibility at the laser frequency is important, and a symmetrical gain curve is expected to yield $\alpha = 0$. Therefore, the biggest contribution to the LEF is expected to be an asymmetry of the gain curve, which in a QCL comes from two main factors; dispersive gain\cite{wacker_coexistence_2007} and non-resonant transitions\cite{jungho_kim_theoretical_2004} (which could or could not involve the any of the laser states). The former effect is due to optical transitions between initial and final states with different in-plane momenta, which can occur as long as the scattering rate is faster than the stimulated emission rate.\cite{willenberg_intersubband_2003} It is most prominent when the initial and final state populations are similar, i.~e.~close to gain clamping high above threshold, and can have a pronounced effect on the LEF of a two-level system\cite{silvestri_coherent_2020, opacak_theory_2019, opacak_frequency_2021}. On the other hand, non-resonant transitions drastically change the shape of the gain curve as well as the dispersion $\beta(\omega)$. In addition, a change in optical intensity redistributes the carriers among all levels, such that $\Delta g$ and $\Delta \beta$ no longer have trivial dependencies on the intensity. For example, $\beta = 0$ does not imply $\alpha = 0$ as in the two-level case, as explained below. In addition, there are multiple other factors that contribute to $\alpha$ in a QCL, such as nonparabolicity\cite{liu_importance_2013} and many-body effects\cite{pereira_intersubband_2011}. The counter-rotating terms in the expression for the susceptibility make a considerable contribution to the LEF near resonance, even for mid-IR transitions.\cite{pereira_intersubband_2011} Finally, effects that are usually overlooked are the intensity-dependent broadening, dipole moment, and energy of the transition, of which we find the latter contribute significantly to the LEF of the studied QCLs.

\begin{figure}
    \centering
    \includegraphics[width=0.8\linewidth]{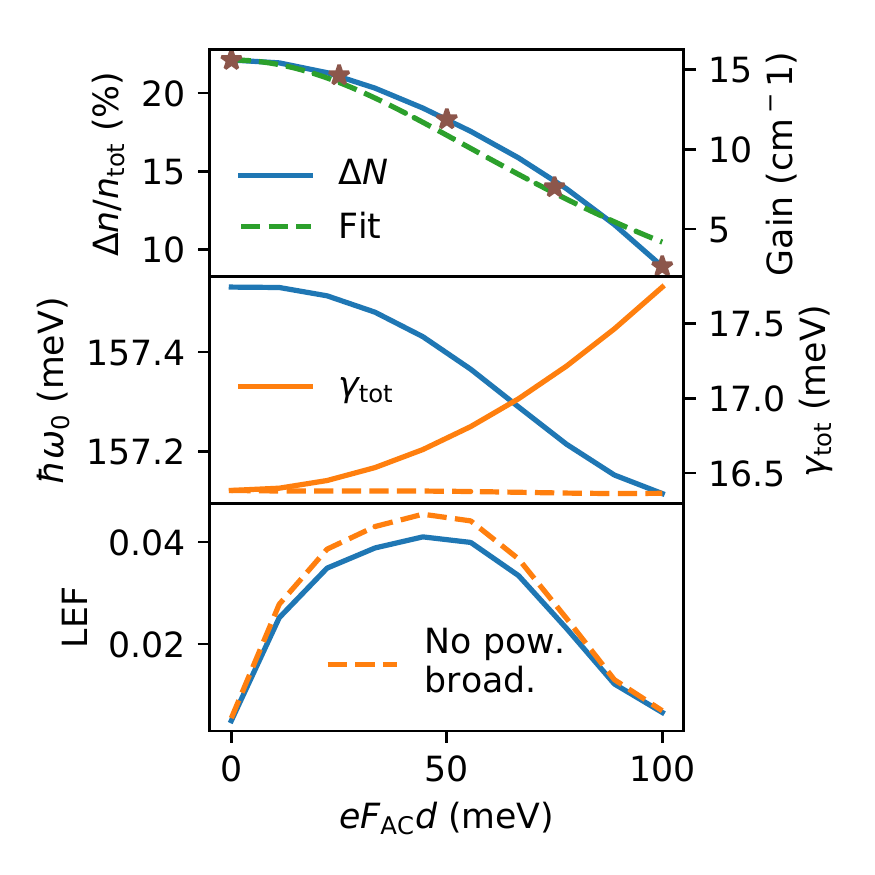}
    \caption{Intensity-dependence of the parameters for the lasing transition which appear in Eq.~\eqref{eq:alpha_diff}. The parameters have been extracted from the NEGF simulations, using a translation into the Wannier-Stark basis. The inversion is shown as a ratio to the total density on both laser states $n_\text{tot} = n_u + n_l$. The total FWHM $\gamma_\text{tot} = (\gamma_\text{sp}^2 + \Omega^2/2 + \gamma_\text{scatt.}^2)^{1/2}$, where broadening due to spontaneous emission $\gamma_\text{sp}$ ($\approx 6\cdot10^{-5}$ meV) and power broadening via the Rabi frequency $\Omega$ have been calculated from these parameters according to Ref.~\cite{loudon_quantum_2000}, and $\gamma_\text{scatt.}(\approx 16$ meV) is the broadening only due to scattering. The decrease in the transition energy $\hbar\omega_0$ as a result of the redistribution of charges, compensated by the increased power broadening due to stimulated emission, results in a non-zero LEF. The green dashed line shows a fit to Eq.~\eqref{eq:saturation}, while the orange dashed lines shows $\gamma_\text{scatt.}$.}
    \label{fig:RWA_resonance}
\end{figure}

By comparing the results from the NEGF model to calculations based on the density matrix formalism (see Appendix \ref{sec:chi_derivation}), we can control for each physical process influencing the LEF. In these calculations, all density-matrix variables are extracted from the NEGF simulations at the respective bias, AC field, and frequency. 
In the simplest case we adopt the rotating wave approximation (RWA), neglect all non-resonant transitions and use a two-level Drude-Lorenz model to find the LEF at resonance (see Eq.~\ref{eq:alpha_diff_detailed}):
\begin{eqnarray}
\alpha^\text{RWA}(\omega=\omega_0)
= \frac{\frac{1}{\gamma}\frac{\d \omega_0}{\d I}}
{\frac{1}{\Delta N}\frac{\d (\Delta N)}{\d I}
+ \frac{2}{z_{ij}}\frac{\d z_{ij}}{\d I}
- \frac{1}{\gamma}\frac{\d \gamma}{\d I}}.
\label{eq:alpha_diff}
\end{eqnarray}
The transition energy $\hbar\omega_0$ is usually assumed to be intensity-independent, with the conclusion that $\alpha = 0$ at the center of the (symmetric) gain curve where $\text{Re}\{\chi(\omega = \omega_0)\} = 0$. Eq.~\eqref{eq:alpha_diff} shows that this is not the case when $\omega_0$ changes as a result of intensity fluctuations, due to e.~g.~a change in mean field potential. This is actually expected from a diagonal transition, which is usually employed in bound-to-continuum designs. Additionally, as shown in Fig.~\ref{fig:RWA_resonance}, although the gain varies in close relation to $\Delta N$, it does not follow Eq.~\eqref{eq:saturation} very well. Already in this simplest case, $\alpha$ can be as large as 0.04, or 10\% of the value of the full simulations, which shows the importance of including at least this effect in analyses of the LEF for mid-IR QCLs.

\begin{figure}
    \centering
    \includegraphics[width=\linewidth]{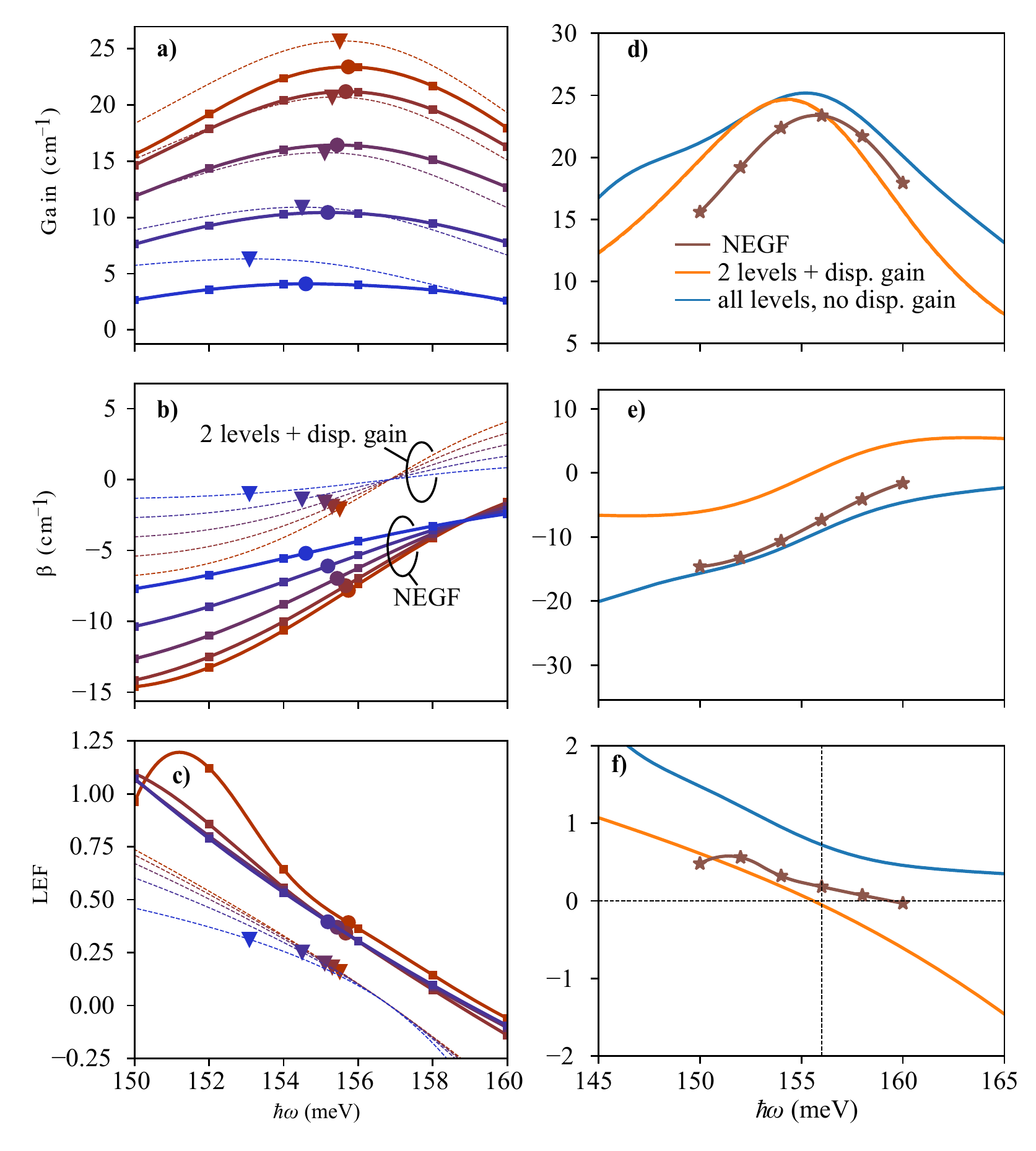}
    \caption{Gain, $\beta$, and LEF as functions of photon energy $\hbar\omega$ at a bias of $eFd = 280$ mV. (a-c) NEGF simulations for different AC field strengths (solid lines), showing a red shift of the gain peak and a small variation of LEF around the peak gain from zero AC field to beyond the gain clamping condition. The dashed lines show the dispersive gain calculations for the reduced two-level system for decreasing inversion. (d-f) 
    Comparison of NEGF to density matrix calculations within different approximations for unsaturated transitions. The overall agreement for gain and $\beta$ as well as the qualitative agreement for the LEF is best when all levels, counter-rotating terms, and no dispersive gain is taken into account. However, significant discrepancy remains with respect to the NEGF calculations which include many-body effects and in-plane distributions. The vertical dashed line in (f) indicates the peak frequency of the NEGF gain curve.
    }
    \label{fig:NEGF_DM}
\end{figure}

Non-resonant transitions modify the susceptibility so that the last equality in Eq.~\eqref{eq:alpha_diff} does not hold. This is illustrated in Fig.~\ref{fig:NEGF_DM} (b), where a good agreement with the NEGF model for $\beta$ is only found when non-resonant transitions are included; even then, a significant discrepancy remains for the line-width enhancement factor.

Adding the dispersive gain (see Eq.~\eqref{eq:dispersive_chi_RW}-\eqref{eq:dispersive_chi_CRW}), the peak gain increases and red shifts. Considering only the upper and lower laser states, this indeed leads to a finite LEF at the peak of the gain curve (see Figs.~\ref{fig:NEGF_DM} c and f). However, adding non-resonant transitions the gain becomes higher and $\beta$ shifts in the opposite direction as compared to the NEGF simulations (see Appendix \ref{sec:details}). This results in a worse agreement for the LEF than for a single transition or non-dispersive gain calculations (it even predicts the wrong sign of the LEF at the peak of the gain curve). The reasons for this inaccuracy could be the assumption of constant subband temperatures and  scattering rates, even for states with large separation in energy and space (although this should be compensated by the small dipole moment $z_{12}$).

Regarding counter-rotating terms, we can estimate their contribution by neglecting all factors but the change in inversion. In this case, we obtain
\begin{equation}
    \alpha^\text{CR}(\omega = \omega_0) = \frac{\mathcal{R}\{\chi^\text{CR}\}}
    {
\mathcal{I}\{\chi^\text{CR}\}
}
= -\frac{\gamma}{2\omega_0} \approx -0.05,
\end{equation}
i.~e.~a small and negative contribution (assuming $\gamma \approx 0.1\omega_0$). This can be seen in Fig.~\ref{fig:NEGF_DM} b) for the two-level approximation. However, non-resonant transitions give a large error to the RWA, since the counter-rotating terms for those transitions are of similar magnitude as the rotating ones. In contrast, for the dispersive gain contributions, since they are inaccurate for the non-resonant transitions in the employed approximations (average transition rates $\gamma = (\gamma_2 + \gamma_1)/2$ for all transitions, and constant subband temperatures of $T = 450$ K), the counter-rotating terms contributes with additional errors, resulting in the LEF actually agreeing slightly better under the RWA. It is therefore recommended to either neglect dispersive gain and take all transitions into account, or to employ a more elaborate treatment of the dispersive gain terms, e.~g.~including the actual subband occupation functions $f_i(E_k)$ as well as decoherence, dephasing, and transition rates.

In conclusion, a wide range of effects contribute to the LEF in an intersubband semiconductor laser, each of which can have significant individual contributions. However, we find that the largest contributions come from non-resonant transitions, dispersive gain, counter-rotating terms, and intensity-dependent transition energy. Other effects, such as nonparabolicity (which has been included in all simulations), are also known to contribute significantly. This implies, that all these effects have to be accounted for when simulating the LEF for a general structure. In addition, comparing density matrix calculations which include all of these effects, to NEGF simulations which include additional many-body effects and $k$-resolved subband distributions, we find that a significant discrepancy to the full NEGF model results remains. This  sensitivity of the LEF to a multitude of complex and inter-related quantum effects, highlights the advantage of the NEGF model for accurate prediction of optical nonlinearities in semiconductor devices, crucial for the development of versatile mid-infrared frequency comb sources. 
\\\\
\section{Acknowledgements}
Financial support from the Qombs project funded by the European Union’s Horizon 2020 research and innovation programme under grant agreement no. 820419 is acknowledged. The simulations were carried out on the Euler computer cluster of ETH Zürich.

\section{Author contributions}
MF performed the NEGF simulations and density-matrix calculations, wrote the manuscript, and made the theoretical derivations. MB conducted the experimental measurements. JF supervised the work.

\def\bibfile{0}

\if\bibfile1

\bibliography{references}

\else

\fi

\begin{appendix}
\begin{widetext}

\section{Derivation of complex conductivity with dispersive gain}
\label{sec:chi_derivation}

Complementing the expressions for the absorption in Ref.~\cite{willenberg_intersubband_2003} with the phases $\beta = \frac{\text{Im}\{\sigma\}}{cn_r\varepsilon_0} = \beta^{qc} + \beta^{bo}$:

\begin{eqnarray}
\beta^{qc} &=& \frac{|z_{21}|^2\omega_0^2}{\varepsilon_0cn_r\omega} \sum_k (f_k^{22} - f_k^{11})\left[ \frac{\varepsilon + \hbar\omega}{(\varepsilon + \hbar\omega)^2 + 4\gamma_k^2} + \frac{\varepsilon - \hbar\omega}{(\varepsilon - \hbar\omega)^2 + 4\gamma_k^2} \right] \\
\beta^{bo} &=& \frac{|z_{21}|^2\omega_0^2}{\varepsilon_0cn_r\omega} \sum_k 2\gamma_k^2
\left[ \frac{(f_{l-}^{22} - f_k^{22}) - (f_{l+}^{11} - f_k^{11})}{(\varepsilon + \hbar\omega)((\varepsilon + \hbar\omega)^2 + 4\gamma_k^2)} +
\frac{(f_{k-}^{22} - f_k^{22}) - (f_{k+}^{11} - f_k^{11})}{(\varepsilon - \hbar\omega)((\varepsilon - \hbar\omega)^2 + 4\gamma_k^2)}\right],
\end{eqnarray}
we will now write down the complex susceptibility for thermalised subbands. Assuming equal scattering rates $\gamma_k = \gamma$, the sum over $k$ can be carried out as for the absorption coefficients, and only the pre-factors are changed. If $\delta \equiv \varepsilon - \hbar\omega > 0$,
\begin{eqnarray}
    &&\sum_k (f_{k-}^{22} - f_k^{22} - f_{k+}^{11}+f_{k}^{11}) = \frac{A}{2\pi}\frac{m^\ast}{\hbar^2}\left( \underbrace{\int_\delta^\infty dE_k e^{-(E_k-\delta-E_F^2)/k_BT}}_{\rightarrow n_2} - \underbrace{\int_\delta^\infty dE_k e^{-(E_k-E_F^2)/k_BT}}_{\rightarrow n_2e^{-\delta/k_BT}} \right) \\
    &+& \frac{A}{2\pi}\frac{m^\ast}{\hbar^2}\left( \underbrace{\int_0^\infty dE_k e^{-(E_k-E_F^1)/k_BT}}_{\rightarrow n_1} - \underbrace{\int_{0}^\infty dE_k e^{-(E_k+\delta-E_F^1)/k_BT}}_{\rightarrow n_1e^{-\delta/k_BT}} \right) \\
    &=& \text{sign}(\delta) (n_1+n_2)(1 - e^{-|\delta|/k_BT})
\end{eqnarray}
and the same result is found if $\delta < 0$. For the non-resonant terms with $\delta' = \varepsilon + \hbar\omega>0$, similarly
\begin{eqnarray}
    &&\sum_k (f_{l-}^{22} - f_k^{22} - f_{l+}^{11}+f_{k}^{11}) = \frac{A}{2\pi}\frac{m^\ast}{\hbar^2}\left( \underbrace{\int_{\delta'}^\infty dE_k e^{-(E_k-\delta'-E_F^2)/k_BT}}_{\rightarrow n_2} - \underbrace{\int_{\delta'}^\infty dE_k e^{-(E_k-E_F^2)/k_BT}}_{\rightarrow n_2e^{-\delta'/k_BT}} \right) \\
    &+& \frac{A}{2\pi}\frac{m^\ast}{\hbar^2}\left( \underbrace{\int_0^\infty dE_k e^{-(E_k-E_F^1)/k_BT}}_{\rightarrow n_1} - \underbrace{\int_{0}^\infty dE_k e^{-(E_k+\delta-E_F^1)/k_BT}}_{\rightarrow n_1e^{-\delta'/k_BT}} \right) \\
    &=& (n_1+n_2)(1 - e^{-\delta'/k_BT}),
\end{eqnarray}
and finally
\begin{equation}
    \sum_k (f_{k}^{22} - f_k^{11}) = n_2 - n_1.
\end{equation}
Therfore, we obtain the complex susceptibility including dispersive gain as
\begin{eqnarray}
    \chi_{12}(\omega) = \frac{|z_{21}|^2\omega_0^2}{\varepsilon_0cn_r\omega^2} \Big{(}
    &\frac{\Delta N(\delta + 2\mathrm{i}\gamma) + (\mathrm{i}\gamma/2|\delta|^{-1} -2\gamma^2) (n_1 + n_2)(1 - e^{-|\delta|/k_BT}) }{\delta^2 + 4\gamma^2}&
    \label{eq:dispersive_chi_RW}
    \\
    + &\frac{\Delta N(\delta' - 2\mathrm{i}\gamma) - (\mathrm{i}2\gamma/2\delta'^{-1} + 2\gamma^2)(n_1 + n_2)(1 - e^{-\delta'/k_BT})}{\delta'^2 + 4\gamma^2}&\Big{)}, \label{eq:dispersive_chi_CRW}
\end{eqnarray}
where Eq.~\eqref{eq:dispersive_chi_RW} accounts for terms in the RWA, and Eq.~\eqref{eq:dispersive_chi_CRW} include the counter-rotating terms. This expression agrees with the corresponding terms given in Refs.~\cite{willenberg_intersubband_2003} and \cite{loudon_quantum_2000}, and is equivalent with the susceptibility given in Ref.~\cite{opacak_frequency_2021} with the re-definitions $\chi \rightarrow \chi^\ast$ and $\alpha \rightarrow -\alpha$.

\end{widetext}

\section{LEF approximations}

Assuming the intensity-dependence of $\chi$ is mainly due to a change in inversion according to Eq.~\eqref{eq:saturation} and the same saturation intensity for all transitions, we may write
\begin{eqnarray}
    \frac{d \chi}{d I} \approx \sum_\eta f_\eta
    \frac{-\Delta N_\eta}{(1 + I/I_\text{sat}))^2I_\text{sat}}
\end{eqnarray}
where $\eta = (i,j)$ is the transition index and $f_\eta \equiv \chi_\eta/\Delta N_\eta$. Thus, the LEF becomes
\begin{eqnarray}
    \alpha \approx \frac{\text{Re}\{\sum_\eta\chi_\eta\}}{\text{Im}\{\sum_\eta\chi_\eta\}} = \frac{\text{Re}\{\chi\}}{\text{Im}\{\chi\}} = -2\frac{\beta}{g}.
\end{eqnarray}

In the RWA approximation at resonanace, we get using the chain rule:
\begin{widetext}
\begin{eqnarray}
\alpha^\text{RWA}(\omega=\omega_0) = \frac{\mathcal{R}\{\Delta \chi\}}{\mathcal{I}\{\Delta\chi\}} = 
\frac{\mathcal{R}\{\chi\}(\frac{1}{\Delta N}\frac{\d (\Delta N)}{\d I}
+ \frac{2}{z_{ij}}\frac{\d z_{ij}}{\d I}
- \frac{1}{\gamma}\frac{\d \gamma}{\d I}) + 
\mathcal{I}\{\chi\}\frac{1}{\gamma}\frac{\d \omega_0}{\d I}}{
\mathcal{I}\{\chi\}(\frac{1}{\Delta N}\frac{\d (\Delta N)}{\d I}
+ \frac{2}{z_{ij}}\frac{\d z_{ij}}{\d I}
- \frac{1}{\gamma}\frac{\d \gamma}{\d I}) -
\mathcal{R}\{\chi\}\frac{1}{\gamma}\frac{\d \omega_0}{\d I}
}
= \frac{\frac{1}{\gamma}\frac{\d \omega_0}{\d I}}
{\frac{1}{\Delta N}\frac{\d (\Delta N)}{\d I}
+ \frac{2}{z_{ij}}\frac{\d z_{ij}}{\d I}
- \frac{1}{\gamma}\frac{\d \gamma}{\d I}}.
\label{eq:alpha_diff_detailed}
\end{eqnarray}
\end{widetext}

\section{More detailed comparison of approximations}
\label{sec:details}

For clarity, Fig.~\ref{fig:NEGF_DM} of the main text shows only two different approximations to the susceptibility. Here, we show additionally the effects of the RWA and non-resonant transitions. We see that the RWA becomes severely flawed when non-resonant transitions are included. We also see that the dispersive gain calculations become inaccurate, even predicting the opposite sign of the LEF, when non-resonant transitions are included. This calls for a more careful treatment of dispersive gain, computing the in-plane distributions and scattering matrix elements, which would add considerable amount of computational load to simplified density-matrix models.

\begin{figure}[h]
    \centering
    \includegraphics[width=0.75\linewidth]{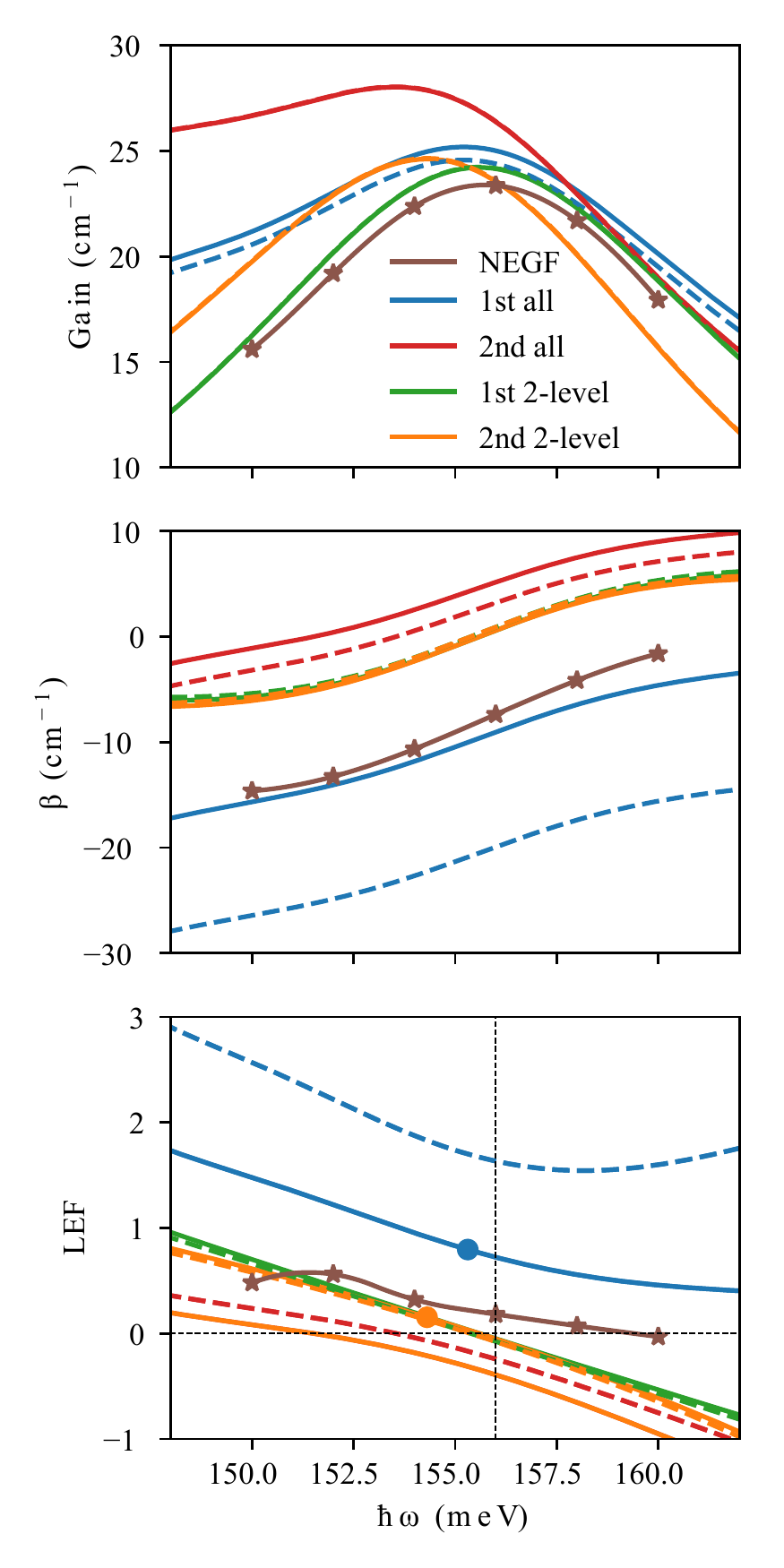}
    \caption{Same as Fig.~\ref{fig:NEGF_DM} (d)-(f) with additional data. Also shown here are results for the RWA (dahsed lines) and dispersive gain with all transitions (2nd all) and non-dispersive gain with two levels (1st 2-level).}
    \label{fig:NEGF_DM_comp_suppl}
\end{figure}
\end{appendix}

\end{document}